\documentclass[iop]{emulateapj}      
\usepackage{courier,graphicx,epsfig,natbib,url,twoopt,color}
\usepackage{hyperref}                
\usepackage{pdfcomment,acronym}      
\usepackage[all]{hypcap}             
\hypersetup{
  colorlinks=true,  
  urlcolor=blue,    
  linkcolor=red     
}
\newdateformat{mydate}{\THEYEAR,\monthname[\THEMONTH], \THEDAy}

\makeatletter
\long\def\frontmatter@title@above{
  \vspace*{-\headsep}\vspace*{\headheight}
   Accepted by {\sc The Astrophysical Journal} 
  \par\vspace*{-\baselineskip}\vspace{12mm}
  }
\def\subtitle{}
\makeatother

\def\revpar{}                 
\long\def\rev#1{#1}           

\def\revpartwo{}                 
\long\def\revtwo#1{#1}           

\def\# #1\par{\par\mbox{}\\ \noindent{\color{red}\small $\sharp$ #1}\\} 

\bibpunct{(}{)}{;}{a}{}{,}    
\makeatletter
 \newcommandtwoopt{\citeads}[3][][]{%
   \nonstopmode
   \href{http://adsabs.harvard.edu/abs/#3}%
        {\def\hyper@linkstart##1##2{}%
         \let\hyper@linkend\@empty\citealp[#1][#2]{#3}}
   \biblink{#3}{\href{http://adsabs.harvard.edu/abs/#3}{ADS}}%
   \errorstopmode}            
 \newcommandtwoopt{\citepads}[3][][]{%
   \nonstopmode
   \href{http://adsabs.harvard.edu/abs/#3}%
        {\def\hyper@linkstart##1##2{}%
         \let\hyper@linkend\@empty\citep[#1][#2]{#3}}
   \biblink{#3}{\href{http://adsabs.harvard.edu/abs/#3}{ADS}}%
   \errorstopmode}            
 \newcommandtwoopt{\citetads}[3][][]{%
   \nonstopmode
   \href{http://adsabs.harvard.edu/abs/#3}%
        {\def\hyper@linkstart##1##2{}%
         \let\hyper@linkend\@empty\citet[#1][#2]{#3}}
   \biblink{#3}{\href{http://adsabs.harvard.edu/abs/#3}{ADS}}%
   \errorstopmode}            
 \newcommandtwoopt{\citeyearads}[3][][]{%
   \nonstopmode
   \href{http://adsabs.harvard.edu/abs/#3}%
        {\def\hyper@linkstart##1##2{}%
         \let\hyper@linkend\@empty\citeyear[#1][#2]{#3}}
   \biblink{#3}{\href{http://adsabs.harvard.edu/abs/#3}{ADS}}%
   \errorstopmode}            
\makeatother

\newcount\longrefs
\def\aap{\ifnum\longrefs=1 {Astron.\ Astrophys.}\else 
                           {A\hbox{\rm \&}A}\fi}
\def\aapr{\ifnum\longrefs=1 {Astron.\ Astrophys.\ Rev.}\else 
                            {A\hbox{\rm \&}AR}\fi}
\def\aaps{\ifnum\longrefs=1 {Astron.\ Astrophys.\ Suppl.}\else 
                            {A\hbox{\rm \&}A Suppl.}\fi}
\def\actaa{\ifnum\longrefs=1 {Acta Astronomica}\else
                            {Acta Astron.}\fi}
\def\aipcs{\ifnum\longrefs=1 {Am.\ Inst.\ Phys.\ Conf.\ Series}\else
                             {AIP Conf.\ Ser.}\fi}
\def\aj{\ifnum\longrefs=1 {Astron.\ J.}\else 
                          {AJ}\fi} 
\def\ao{\ifnum\longrefs=1 {Applied Optics}\else 
                           {Appl.\ Opt.}\fi} 
\def\aspcs{\ifnum\longrefs=1 {Astron.\ Soc.\ Pacific Conf.\ Series}\else 
                           {ASP Conf.\ Ser.}\fi} 
\def\apj{\ifnum\longrefs=1 {Astrophys.\ J.}\else 
                           {ApJ}\fi} 
\def\apjl{\ifnum\longrefs=1 {Astrophys.\ J. Lett.}\else 
                            {ApJL}\fi} 
\def\aplett{\ifnum\longrefs=1 {Astrophys.\ J. Lett.}\else 
                            {ApJ}\fi} 
\def\apjs{\ifnum\longrefs=1 {Astrophys.\ J. Suppl.}\else 
                            {ApJS}\fi}
\def\apss{\ifnum\longrefs=1 {Astrophys.\ and Space Science}\else 
                            {Astrophys.\ Space Sci.}\fi}
\def\araa{\ifnum\longrefs=1 {Ann.\ Rev.\ Astron.\ Astrophys.}\else 
                            {ARA\hbox{\rm \&}A}\fi}
\def\azh{\ifnum\longrefs=1 {Astronomicheskii Zhurnal}\else 
                            {Astron.\ Zhur.}\fi}
\def\baas{\ifnum\longrefs=1 {Bull.\ Am.\ Astron.\ Soc.}\else 
                            {BAAS}\fi}
\def\bain{\ifnum\longrefs=1 {Bull.\ Astronom.\ Institutes Netherlands}\else
                            {Bull.\ Astr.\ Inst.\ Neth.}\fi}
\def\cjaa{\ifnum\longrefs=1 {Chinese Jour.\ Astron.\ Astrophys.}\else 
                            {Chin.\ J.\ A\&A}\fi}
\def\gca{\ifnum\longrefs=1 {Geochim.\ Cosmochim.\ Acta}\else 
                           {Geochim.\ Cosmochim.\ Acta}\fi}
\def\grl{\ifnum\longrefs=1 {Geophys.\ Res.\ Lett.}\else 
                           {Geoph.\ Res.\ Lett.}\fi}
\def\iaucirc{\ifnum\longrefs=1 {IAU Circulars}\else 
                          {IAU Circ.}\fi}
\def\icarus{\ifnum\longrefs=1 {Icarus}\else 
                          {Icarus}\fi}
\def\ip{\ifnum\longrefs=1 {in press}\else 
                          {in press}\fi}
\def\jcap{\ifnum\longrefs=1 {Jour.\ Cosmology Astropart.\ Phys.}\else 
                          {JCAP}\fi}
\def\jgr{\ifnum\longrefs=1 {J.\ Geophys.\ Res.}\else 
                           {J.\ Geophys.\ Res.}\fi}  
\def\jrasc{\ifnum\longrefs=1 {J.\ Royal Astron.\ Soc.\ Canada}\else 
                           {JRAS Can.}\fi}  
\def\memsai{\ifnum\longrefs=1 {Mem.~Soc.~Astron.~Italiana}\else
                              {MmSAI}\fi}
\def\mnras{\ifnum\longrefs=1 {Mon.\ Not.\ Roy.\ Astron.\ Soc.}\else 
                             {MNRAS}\fi} 
\def\na{\ifnum\longrefs=1 {New Astronomy}\else 
                           {New Astron.}\fi}
\def\nar{\ifnum\longrefs=1 {New Astronomy rev.}\else 
                           {New Astron.\ Rev.}\fi}
\def\nat{\ifnum\longrefs=1 {Nature}\else 
                           {Nat}\fi}
\def\pasa{\ifnum\longrefs=1 {Pub.\ Astron.\ Soc.\ Australia}\else 
                            {PASA}\fi} 
\def\pasj{\ifnum\longrefs=1 {Pub.\ Astron.\ Soc.\ Japan}\else 
                            {PASJ}\fi} 
\def\pasp{\ifnum\longrefs=1 {Pub.\ Astron.\ Soc.\ Pacific}\else 
                            {PASP}\fi} 
\def\physscr{\ifnum\longrefs=1 {Physica Scripta}\else 
                            {Phys.\ Scrip.}\fi} 
\def\planss{\ifnum\longrefs=1 {Planetary \& Space Science}\else 
                            {Plan. \& Space Sci.}\fi} 
\def\procspie{\ifnum\longrefs=1 {Proc.\ SPIE}\else 
                            {Proc.\ SPIE}\fi} 
\def\qjras{\ifnum\longrefs=1 {Quarterly J.\ Royal Astron.\ Soc.}\else 
                            {QJRAS}\fi} 
\def\rmxaa{\ifnum\longrefs=1 {Revista Mexicana de Astron.\ y Astrofys.}\else 
                            {RMxAA}\fi} 
\def\sa{\ifnum\longrefs=1 {Soviet Astron..}\else 
                               {Sov.\ Astron.}\fi}
\def\skytel{\ifnum\longrefs=1 {Sky \& Telescope}\else 
                            {Sky \& Tel.}\fi} 
\def\solphys{\ifnum\longrefs=1 {Solar Phys.}\else 
                               {SoPh}\fi}
\def\sovast{\ifnum\longrefs=1 {Soviet Astronomy}\else 
                               {Sov.\ Ast.}\fi}
\def\ssr{\ifnum\longrefs=1 {Space Science Rev.}\else 
                               {Space\ Sci.\ Rev.}\fi}
\def\zap{\ifnum\longrefs=1 {Zeitschr.\ f.\ Astrophysik}\else
                               {Z.\ Astrophys.}\fi}

\makeatletter
\newcommand{\bibnote}[2]{\@namedef{#1note}{#2}}
\newcommand{\biblink}[2]{\@namedef{#1link}{#2}}
\makeatother

\def\acdef#1{\acl{#1} ({#1})}     
\newacro{AA}{Astronomy \& Astrophysics}
\newacro{ADS}{Astrophysics Data System}
\newacro{AIA}{Atmospheric Imaging Assembly}
\newacro{AO}{adaptive optics}
\newacro{ApJ}{Astrophysical Journal}
\newacro{AR}{active region}
\newacro{BFI}{Broad-band Filter Imager}
\newacro{CE}{coronal equilibrium}
\newacro{CfA}{Center for Astrophysics}
\newacro{CME}{coronal mass ejection}
\newacro{CRD}{complete redistribution}
\newacro{CRISP}{CRisp Imaging SpectroPolarimeter}
\newacro{CRISPEX}{CRisp SPectral EXplorer}
\newacro{CS}{coherent scattering}
\newacro{DEM}{Differential Emission Measure}
\newacro{DKIST}{Daniel K. Inouye Solar Telescope}
\newacro{DLR}{Deutsches Zentrum f\"ur Luft- und Raumfahrt}
\newacro{DOT}{Dutch Open Telescope}
\newacro{DST}{Richard B. Dunn Solar Telescope}   
\newacro{EB}{Ellerman bomb}
\newacro{EDP}{\'{E}dition Diffusion Presse Sciences}  
\newacro{EIT}{Extreme ultraviolet Imaging Telescope}
\newacro{EPIC}{European participation in Solar-C}
\newacro{ERC}{European Research Council}
\newacro{ESA}{European Space Agency}
\newacro{EST}{European Solar Telescope}
\newacro{EUV}{extreme ultraviolet}
\newacro{FAF}{flaring arch filament}
\newacro{FITS}{Flexible Image Transport System}
\newacro{FOV}{field of view}
\newacro{fov}{field of view}
\newacro{FWHM}{full width at half maximum}
\newacro{HAO}{High Altitude Observatory}
\newacro{HD}{hydrodynamics}
\newacro{Hi-C}{High Resolution Coronal Imager Sounding Rocket}
\newacro{HMI}{Helioseismic and Magnetic Imager}
\newacro{IAA}{Instituto de Astrof\'{i}sica de Andaluc\'{i}a}
\newacro{IAC}{Instituto de Astrof\'{i}sica de Canarias}
\newacro{IAS}{Institut d'Astrophysique Spatiale}
\newacro{IDL}{Interactive Data Language}
\newacro{IMaX}{Imaging Magnetograph eXperiment}
\newacro{INAF}{Istituto Nazionale di Astrofisica}
\newacro{IB}{IRIS bomb}
\newacro{IR}{infrared}
\newacro{IRIS}{Interface Region Imaging Spectrograph}
\newacro{ISAS}{Institute of Space and Astronautical Science}
\newacro{ISP}{Institute for Solar Physics}
\newacro{ISS}{International Space Station}
\newacro{ISSI}{International Space Science Institute}
\newacro{ITA}{Institute for Theoretical Astrophysics}
\newacro{JAXA}{Japan Aerospace Exploration Agency}
\newacro{KIS}{Kiepenheuer--Institut f\"{u}r Sonnenphysik}
\newacro{KPNO}{Kitt Peak National Observatory}
\newacro{LASP}{Laboratory for Atmospheric and Space Physics}
\newacro{LC}{liquid cristal}
\newacro{LMSAL}{Lockheed Martin Solar and Astrophysics Labratory}
\newacro{LOS}{line of sight}
\newacro{LTE}{local thermodynamic equilibrium}
\newacro{MC}{magnetic concentration}
\newacro{MCAO}{multi-conjugate adaptive optics} 
\newacro{MDI}{Michelson Doppler Imager}
\newacro{ME}{Milne-Eddington} 
\newacro{MHD}{magnetohydrodynamics}
\newacro{MOMFBD}{Multi-Object Multi-Frame Blind Deconvolution}
\newacro{MPE}{Max--Planck--Institut f\"ur extraterrestrische Physik}
\newacro{MPG}{Max--Planck--Gesellschaft}
\newacro{MPS}{Max Planck Institute for Solar System Research}
\newacro{MSSL}{Mullard Space Science Laboratory}
\newacro{MTF}{modulation transfer function}
\newacro{NAOJ}{National Astronomical Observatory of Japan}
\newacro{NASA}{National Aeronautics and Space Administration}
\newacro{NLTE}{non-local thermodynamic equilibrium}
\newacro{NOAA}{National Oceanic and Atmospheric Administration}
\newacro{non-E}{non-equilibrium}
\newacro{NSO}{National Solar Observatory}
\newacro{NWO}{Netherlands Organisation for Scientific Research}
\newacro{PRD}{partial redistribution}
\newacro{PROBA2}{PRoject for OnBoard Autonomy}
\newacro{PSF}{point spread function}
\newacro{QS}{quiet Sun}
\newacro{RAL}{Rutherford Appleton Laboratory}
\newacro{R-MHD}{radiation hydrodynamics}
\newacro{rms}{root mean square}
\newacro{RMS}{root mean square}
\newacro{ROB}{Royal Observatory of Belgium}
\newacro{ROI}{region of interest}
\newacro{RTE}{radiative transfer equation}
\newacro{SE}{statistical equilibrium}
\newacro{SDO}{Solar Dynamics Observatory}
\newacro{SJI}{slit-jaw image}
\newacro{SNR}{signal-to-noise ratio}
\newacro{SO}{Solar Orbiter}
\newacro{SoHO}{Solar and Heliospheric Observatory}
\newacro{SP}{Spectropolarimeter}
\newacro{SST}{Swedish 1-m Solar Telescope}
\newacro{SUMER}{Solar Ultraviolet Measurements of Emitted Radiation}
\newacro{SUFI}{Sunrise Filter Imager}
\newacro{SVD}{singular value decomposition}
\newacro{SVST}{Swedish Vacuum Solar Telescope}
\newacro{THEMIS}{T\'{e}lescope H\'{e}liographique pour l'Etude du 
   Magn\'{e}tisme et des Instabilit\'{e} Solaires}     
\newacro{TR}{transition region}
\newacro{TRACE}{Transition Region and Coronal Explorer}
\newacro{TSI}{total solar irradiance}
\newacro{UV}{ultraviolet}
\newacro{VIRGO}{Variability of solar IRradiance and Gravity Oscillations}
\newacro{VTT}{Vacuum Tower Telescope}    
\newacro{XRT}{X-Ray Telescope}

\def\acp#1{\pdftooltip{\acs{#1}}{\acl{#1}}}  

\def\nl{,\ } 

\def\ITA{Institute of Theoretical Astrophysics\nl
         University of Oslo\nl
         P.O. Box 1029, Blindern\nl N-0315 Oslo\nl Norway}

\def\LA{Lingezicht Astrophysics\nl 't Oosteneind 9\nl 4158\,CA Deil\nl 
        The Netherlands}

\long\def\startignore #1\stopignore{}   

\def\rmit#1{{\it #1}}              
           
\def\ie{\rmit{i.e.,}}              

\def\specchar#1{\uppercase{#1}}    
\def\specand{\,\&\,}               
\def\specand{ and }                

\def\CII{\mbox{C\,\specchar{ii}}} 
 
\def\CIV{\mbox{C\,\specchar{iv}}}

\def\MgI{\mbox{Mg\,\specchar{i}}}

\def\MnI{\mbox{Mn\,\specchar{i}}}

\def\NaI{\mbox{Na\,\specchar{i}}}

\def\SiIV{\mbox{Si\,\specchar{iv}}}

\def\Halpha{\mbox{H\hspace{0.1ex}$\alpha$}} 

\def\NaID{\mbox{Na\,\specchar{i}\,\,D}}
\def\NaIDone{\mbox{Na\,\specchar{i}\,\,D$_1$}}

\def\MgIb{\mbox{Mg\,\specchar{i}\,b}}

\def\MgIbtwo{\mbox{Mg\,\specchar{i}\,b$_2$}}

\def\CaIIH{\mbox{Ca\,\specchar{ii}\,\,H}}
\def\CaIIHK{\mbox{Ca\,\specchar{ii}\,\,H{\specand}K}}

\def\MgIIhk{\mbox{Mg\,\specchar{ii}{\specand}k}}

\def\level #1 #2#3#4{$#1 \; ^{#2} \mbox{#3} ^{#4}$}   

\def\deg{\hbox{$^\circ$}}       

\def\is{\!=\!}                             
\def\={\hbox{$\!=\!$}}                     

\def\rmit#1{#1}                 
\def\specchar#1{{\textsc{#1}}}  

\def\deffigs{deffigs}
\def\deffigs{}

\begin{document}

\title{Ellerman bombs at high resolution. 
IV. 
Visibility in \NaI\ and \MgI} \subtitle{}

\author{R. J. Rutten\altaffilmark{1}}
\author{L. H. M. Rouppe van der Voort}
\author{G. J. M. Vissers}

\affil{\ITA} 
\altaffiltext{1}{Address: \LA. 
Email: R.J.Rutten@uu.nl.
Website: \url{http://www.staff.science.uu.nl/~rutte101}.}

\shorttitle{Ellerman bombs in \NaI\ and \MgI}
\shortauthors{Rutten et al.}

\begin{abstract}
Ellerman bombs are transient brightenings of the wings of the solar
Balmer lines that mark reconnection in the photosphere.
Ellerman noted in 1917 that he did not observe such brightenings in
the \NaID\ and \MgIb\ lines.
\rev{This non-visibility should constrain EB interpretation, but has
not been addressed in published bomb modeling.
We therefore test Ellerman's observation and confirm it} using
high-quality imaging spectrometry with the Swedish 1-m Solar
Telescope.
However, we find diffuse brightness in these lines that seems to
result from prior EBs.
We \rev{tentatively} suggest this is a post-bomb hot-cloud phenomenon
also found in recent EB spectroscopy in the ultraviolet.
\end{abstract}


\section{Introduction}\label{sec:introduction}
\citetads{1917ApJ....46..298E} 
discovered intense short-lived brightenings of the extended wings of
the Balmer \Halpha\ line at 6563\,\AA\ in complex solar active regions
which he called ``solar hydrogen bombs.''
They are called ``Ellerman bombs'' (henceforth abbreviated to EB)
since \citetads{1960PNAS...46..165M} 
but they have also been called ``moustaches'' after 
\citetads{1956Obs....76..241S} 
(who did not call them bombs but did invoke nuclear explosions).
For more detail we refer to our review of the extensive \acp{EB}
literature in \citetads{2013JPhCS.440a2007R}. 

In the preceding \acp{EB} studies of this series we used high-quality
imaging spectroscopy and spectropolarimetry with the \acdef{SST}.
Paper~I (\citeads{2011ApJ...736...71W}) 
established that \acp{EB}s are a purely photospheric phenomenon.
Paper~II (\citeads{2013ApJ...774...32V}) 
added evidence that \acp{EB}s mark magnetic reconnection of strong
opposite-polarity field concentrations in the low photosphere and
discussed their appearance in ultraviolet continuum images from the
\acdef{AIA} of the {\em \acdef{SDO}}.
Paper~III (\citeads{Vissers-etal-submitted-ApJ}) 
discussed the appearance of \acp{EB}s in ultraviolet lines in spectra taken
with the \acdef{IRIS}
\rev{(\citeads{2014SoPh..289.2733D})}. 

  \bibnote{2011ApJ...736...71W}{(Paper~I)}
  \def\PaperI{\href{http://adsabs.harvard.edu/abs/2011ApJ...736...71W}
             {Paper~I}}
  \bibnote{2013ApJ...774...32V}{(Paper~II)}
  \def\PaperII{\href{http://adsabs.harvard.edu/abs/2013ApJ...774...32V}
              {Paper~II}}
  \bibnote{Vissers-etal-submitted-ApJ}{(Paper~III)}
  \def\PaperIII{\href{http://adsabs.harvard.edu/abs/Vissers-et-al-submitted-ApJ}{Paper~III}}

In this installment we address the visibility of
\acp{EB}s in the \NaID\ and \MgIb\ lines.
\citetads{1917ApJ....46..298E} 
already remarked that his bombs did not appear in these lines, nor in
the continuum. 
We check his claims here exploiting the high resolution of the
\acp{SST}
\rev{because both provide important formation constraints for
\acp{EB} interpretation.}

The absence of \acp{EB}s in the continuum was already an important
constraint for the \acp{EB} modeling by
\citetads{1983SoPh...87..135K}, 
\citetads{2010MmSAI..81..646B}, 
\citetads{2013A&A...557A.102B}, 
and
\citetads{2014A&A...567A.110B} 
who all applied ad-hoc perturbations of a static standard model and
NLTE line synthesis to reproduce observed \Halpha\ moustaches.
\rev{We summarize these studies briefly.

\citetads{1983SoPh...87..135K} 
fitted \acp{EB} profiles in \Halpha\ and \CaIIHK\ spectrograms taken at Hida
Observatory by perturbing the VAL3C standard model atmosphere of
\citetads{1981ApJS...45..635V}. 
He concluded that temperature enhancement of 1500\,K and density
enhancement by a factor of 5 were needed at heights 700--1200\,km,
\ie\ in the VAL3C chromosphere, and excluded deeper onsets because
these predicted continuum brightening.

\citetads{2010MmSAI..81..646B} 
modeled \acp{EB} contrasts in Lyot-filter \Halpha\ images from the
Dutch Open Telescope (DOT) and simultaneous ultraviolet images at
1600\,\AA\ from the
\acp{TRACE} satellite
by perturbing a similar but more recent standard model, but only its
temperature.
They concluded that a hump-like increase of about 3000\,K in the upper
photosphere would explain their measurements.

In a similar analysis
\citetads{2014A&A...567A.110B} 
used another \acp{DOT} data set which also provided concurrent \CaIIH\
images and extended the model perturbations to large grids, also
modifying the density. 
Their best fit was for a model only perturbing the temperature, by
4000\,K at height 1000\,km in the onset of the model chromosphere.  

The analysis by
\citetads{2013A&A...557A.102B} 
was the most elaborate, observationally by using Fabry--P\'erot imaging
spectroscopy in \Halpha\ with concurrent full-Stokes polarimetry at
the German Vacuum Tower Telescope, and in modeling by not applying the
1D plane-parallel layer assumption taken by the other authors but
instead performing 2D \Halpha\ synthesis for imposed \acp{EB}
perturbations embedded within a standard model and including slanted
limbward viewing.   
They obtained best fits from temperature increases in the 300--800\,km
height range up to 5000\,K (doubling of their standard model) and
simultaneous increase of the \Halpha\ opacity by a factor of five.

In all these models the proposed temperature enhancements were humps
starting above 300\,km or higher to avoid enhancements} of 
the optical continuum.
However, our high-resolution observations in \PaperI\ contradict such
high \acp{EB} onset because they show that actual \acp{EB}s are rooted
deep in strong-field magnetic concentrations in network lanes, without
apparent gap of such size.
This lower-part \acp{EB} visibility \rev{in \Halpha}\ remains unexplained.

A potential second failure of the models is that none was verified
with respect to Ellerman's non-appearance of \acp{EB}s in the \MgIb\
and \NaID\ lines.
In standard models these lines sample the heights of the imposed
temperature enhancements
(\citeads{2011A&A...531A..17R}), 
so that the \acp{EB} models will probably predict \revpar \acp{EB}
brightening in them.

A third, more recent, issue is the pronounced visibility of \acp{EB}s
in the \CII\ 1334 and 1335\,\AA\ and \SiIV\ 1394 and 1403\,\AA\ doublets
sampled by \acp{IRIS} that we reported in \PaperIII. 
It seems unlikely that the published models can match their
brightenings since they impose \acp{EB} temperatures reaching 
\rev{10,000\,K at most,} whereas the characteristic formation
temperature of the \SiIV\ lines is 80,000\,K \revpar
\rev{(\citeads{2014SoPh..289.2733D})}. 

In this paper we concentrate on the earlier issues: the non-appearance
of \acp{EB}s in the continuum and in the \NaID\ and \MgIb\ lines.
We use high-resolution \acp{SST} observations to confirm Ellerman's
claims, but we do find more diffuse brightenings that seem related to
prior \acp{EB} activity detected in {\em \acp{SDO}}/\acp{AIA} images. 
These may represent hot \acp{EB} aftermaths as those
diagnosed from \acp{IRIS} spectra (\PaperIII).

In Section~\ref{sec:observations} we describe the \acp{SST} observations taken
for this analysis and in Section~\ref{sec:results} the results, including
comparison with {\em \acp{SDO}} images.
The conclusions follow in Section~\ref{sec:conclusion}.

\begin{figure*}
  \centerline{\includegraphics[width=\textwidth]{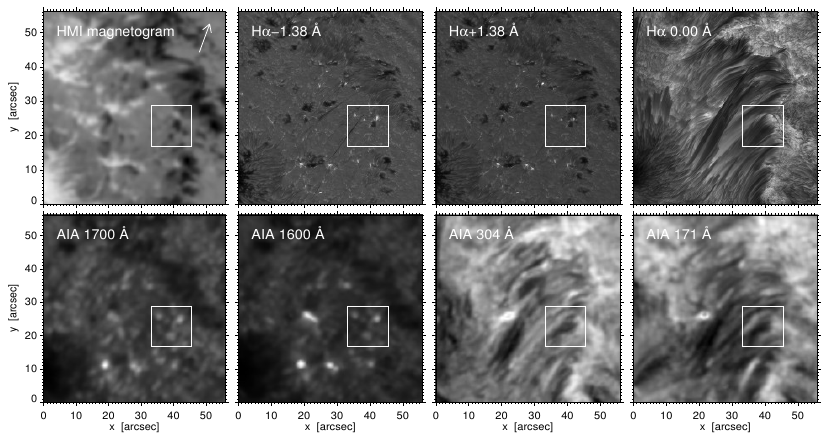}}
  \caption[]{\label{fig:full} 
  Full-field images from \acp{SST}/\acp{CRISP} taken on 2013 July 4, 
  at 11:22:05~UT and simultaneous co-aligned image cutouts from
  {\em \acp{SDO}}. 
  The orientation is not heliographic because the field of view of the
  \acp{SST} rotates with time due to its alt-azimuth heliostat.
  Here it has been rotated 180\deg\ in order to make upright \acp{EB}s
  appear upright; the arrow in the first panel specifies the projected
  local upright direction (to the nearest limb).
  {\em Upper row\/}: \acp{HMI} magnetogram, blue and red wings of
  \Halpha\ at $\Delta \lambda \is \pm 1.38$\,\AA\ from 
  line center, \Halpha\ line center.
  {\em Lower row\/}: 1700, 1600, 304 and 171\,\AA\ images from
  \acp{AIA}, with the first two showing the square root of the
  measured intensity, the other two the logarithm.
  The white frame outlines the small cutout subfield of
  Figures~\ref{fig:cut} and ~\ref{fig:sdo}.
  }
\end{figure*}

\section{Observations}
\label{sec:observations}

2013 July 4
was a day of variable seeing at the \acp{SST}
(\citeads{2003SPIE.4853..341S}) 
on La Palma. 
These conditions and the presence of an emerging active region toward
the limb were suited to test Ellerman's claims through imaging
spectroscopy with the \acl{CRISP} (CRISP;
\citeads{2008ApJ...689L..69S}), 
a profile-scanning Fabry--P\'erot interferometer.
The target was AR\,1785 at solar $(X,Y)\is(-693,-189)$~arcsec from
disk center, with viewing angle $\mu \is 0.65$
($\theta \is 49\fdg5$).

Dual profile scanning of \Halpha\ and the \NaIDone\ line at
5895.94\,\AA\ was performed during 11:09--11:26\,UT, followed by dual
profile scanning of \Halpha\ and the \MgIbtwo\ line at 5172.70\,\AA\
during 11:28--11:41\,UT (wavelengths for standard air from
\citeads{1966sst..book.....M}). 
The scan cadence was 20\,s.

\Halpha\ was sampled at 39 wavelengths \rev{over
$\Delta \lambda \is \pm 2.06$\,\AA\ from the center of the profile
after averaging over the full field of observation}, with equidistant
spacing $\Delta \lambda \is 0.086$\,\AA\ across its core and wider
spacing in its wings. 
\NaIDone\ was sampled at 41 wavelengths over
$\Delta \lambda \is \pm 1.71$\,\AA\ with narrowest core sampling
$\Delta \lambda \is 0.028$\,\AA; \MgIbtwo\ similarly over
$\Delta \lambda \is \pm 0.89$\,\AA.

Movies made from these data showed that the \acp{EB}s present in
\Halpha\ were indeed not evident in \NaIDone\ or \MgIbtwo,
but they did suggest that at the locations where \Halpha\ \acp{EB}s
went off diffuse brightening in these lines occurred subsequently.

Because the image quality was too variable for time-sequence analysis
we selected the sharpest moment of each dual profile scan and
performed full reduction only for these, using procedures described
by \citetads{2015A&A...573A..40D}. 
These include dark- and flat-field correction, multi-object multi-frame
blind deconvolution following
\citetads{2005SoPh..228..191V} 
to reduce the effects of high-order atmospheric seeing not already
corrected by the adaptive optics of the \acp{SST}, removal of
remaining small-scale distortions between the different line
samplings, and correction for the prefilter transmission profile.

We also collected corresponding longer-duration image sequences from
{\em \acp{SDO}}/\acp{AIA} (\citeads{2012SoPh..275...17L}) 
and {\em \acp{SDO}}/\acp{HMI}
(\citeads{2012SoPh..275..207S}) 
using the JSOC image cutout service at Stanford University.
They were co-aligned with the \acp{SST} images using \acp{IDL}
programs available on the website of the first author.

In the alignment and the data analysis we made much use of the
\acl{CRISPEX} (CRISPEX;
\citeads{2012ApJ...750...22V}) 
for data browsing. 

\begin{figure*}
  \centerline{\includegraphics[width=\textwidth]{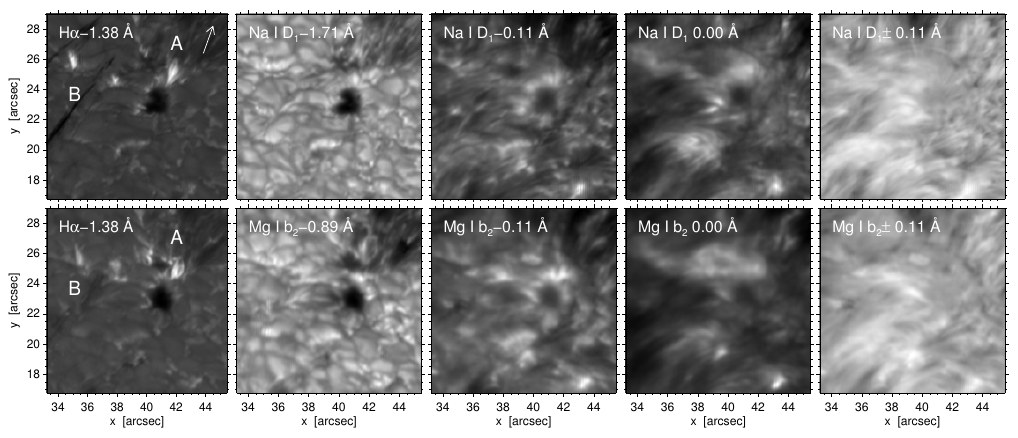}}
  \caption[]{\label{fig:cut} 
  \acp{EB} visibility in the \NaIDone\ and \MgIbtwo\ lines for the
  subfield defined in Figure~\ref{fig:full}.
  {\em Upper row\/}: simultaneous samplings in the outer blue wing of
  \Halpha, the outer blue wing, inner blue wing, and line center of
  the Na line, and an inner-wing Na Dopplergram.
  These were all taken at 11:22:05~UT. 
  {\em Lower row\/}: idem for the Mg line, taken at 11:29:47~UT.
  The wavelength separations from line center are specified in each
  panel.
  The arrow in the first panel specifies the projected upright
  direction.
  The \revtwo{\acp{EB}s below A in the two \Halpha\ panels are}
  enlarged in Figure~\ref{fig:tiny}; \revtwo{those above B in
  Figure~\ref{fig:tiny-b}.}
  }
\end{figure*}

\section{Results}
\label{sec:results}

\paragraph{Overview}
Figure~\ref{fig:full} gives an overview of the observed area
at the time of the best \acp{SST} \Halpha\,-\,\NaIDone\ scan.

The {\em \acp{SDO}}/\acp{HMI} magnetogram in the first panel illustrates the
magnetic complexity of the emerging region, with many adjacent
opposite-polarity field patches.
The next two panels sample the blue and red outer wings of \Halpha\
and contain multiple \acp{EB}s. 
They tend to be located near adjacent opposite-polarity field patches
in the first panel.
However, the sensitivity and resolution of \acp{HMI} are too poor to
resolve the cancelation of small opposite-polarity patches that
characteristically occurs at \acp{EB} sites (\PaperII).

The fourth panel illustrates that \acp{EB}s take place well underneath the
thick fibrilar canopy which the \Halpha\ line core invariably displays
in such crowded active regions. 
The bright feature just left of image center is not an \acp{EB}.

The \acp{AIA}\,1700\,\AA\ image in the bottom row illustrates that
\acp{EB}s are well observable in this diagnostic (\PaperII).
The round bright features show good correspondence with the \Halpha\
\acp{EB}s, be it at inferior spatial resolution.  
In \PaperII\ we showed that the morphology is not identical; in
\PaperIII\ that in slanted viewing the 1700\,\AA\ images show mostly
the downward-moving cooler lower part of an \acp{EB}, whereas the
ultraviolet \acp{IRIS} diagnostics favor the upward-moving hotter
upper part. 
The whole \acp{EB} is seen in the \Halpha\ wings.

In the \acp{AIA}\,1600\,\AA\ image in the second row of
Figure~\ref{fig:full} the \acp{EB}s reach yet larger brightness
contrast, but here the non-\acp{EB} feature left of center in the
\Halpha\ core image appears also very bright.
This is a small flaring arch filament (FAF) of which the brightness is
likely contributed by the \CIV\ doublet at 1548 and 1550\,\AA\ in this
passband. 
Such small \acp{FAF}s typically have elongated shapes, as is the case
here, live shorter, show rapid apparent motion along filamentary
strands, and stand out also in higher-temperature \acp{AIA}
diagnostics (\PaperIII).
The feature is indeed also prominent in the remaining two panels
showing \acp{AIA}'s 304 and 171\,\AA\ images, whereas the \acp{EB}s
leave no signature in these (\PaperII). 

In passing, we note that the darkest \Halpha\ line-core fibrils are
also visible in the hot \acp{AIA} diagnostics. 
The darkest fibrils have the largest neutral-hydrogen density
(\citeads{2012ApJ...749..136L}) 
and hence block short-wavelength radiation by incoherent bound-free
scattering out of the narrow \acp{AIA} passbands
(\citeads{1999ASPC..184..181R}). 

\paragraph{EBs in \NaIDone\ and \MgIbtwo}
Figure~\ref{fig:cut} presents some \acp{EB}s in more detail, including
the scene in \NaIDone\ and \MgIbtwo\ for the subfield defined by the
frame in Figure~\ref{fig:full}.

The first column shows this area in the blue wing of \Halpha.
In limbward viewing at this image quality the distinction between
\acp{EB}s and much more ubiquitous quiescent magnetic concentrations
is easily made.
However, these have often been confused, in the older \acp{EB}
literature (see
\citeads{2013JPhCS.440a2007R}) 
but also more recently (details in \PaperIII). 

Although the rows differ by nearly 8~minutes, a pair of unmistakable and
rather similar \acp{EB}s is seen 
below A 
in the two first-column images.
We would not call this the same \acp{EB} because \acp{EB}s tend to
come in rapid succession, appearing as ``flickering flames'' whose
feet travel along an intergranular lane filled with magnetic
concentrations (\PaperI) canceling against incoming fresh
opposite-polarity flux (\PaperII).
This often happens repetitively in sequences which may last many
minutes and may even repeat during an hour or longer (\PaperIII).
In this case, at eight minutes separation, this pair shows a small
shift in location but roughly similar morphology including upward
splits that suggest reconnection along different field lines or
multiple field-guided jets. 
Both show upright orientation at least in their azimuthal projection. 

\begin{figure*}
  \centerline{\includegraphics[width=0.8\textwidth]{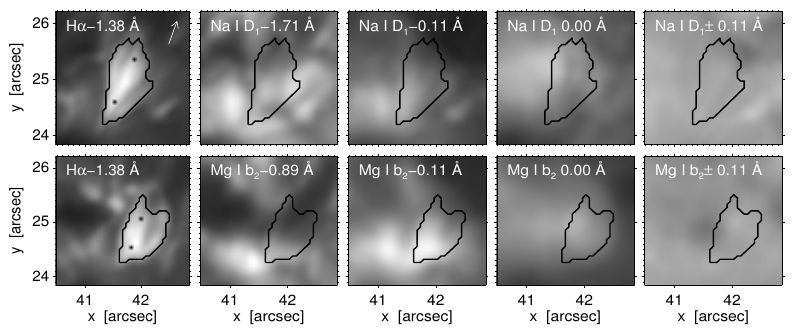}}
  \caption[]{\label{fig:tiny} 
  Magnification of the \acp{EB}s \revtwo{below A in Figure~\ref{fig:cut}}.
  The panel ordering and bytescaling are the same as in
  Figure~\ref{fig:cut}.
  The \acp{EB}s in the first column are outlined by \Halpha\ intensity
  contours that are overlaid in the other columns for reference.
  The black pixels in the first column mark lower and
  upper \acp{EB} sampling locations for Figure~\ref{fig:spectra}.
 }
\end{figure*}

\begin{figure}
  \centerline{\includegraphics[width=\columnwidth]{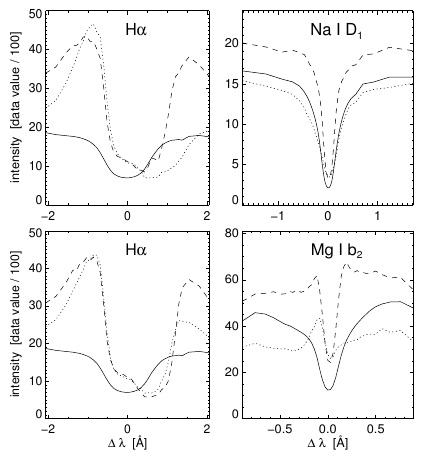}}
  \caption[]{\label{fig:spectra} Spectral profiles at the two sampling
  times.
  {\em Solid\/}: mean over the whole field of view.
  {\em Dashed\/}: lower \acp{EB} pixel specified in the first column
  of Figure~\ref{fig:tiny}.
  {\em Dotted\/}: upper pixel.
  The wavelengths are relative to the line center of the mean
  profiles. 
  The intensities are data readouts divided by 100.
  }
\end{figure}

Smaller \acp{EB}s are seen 
above B
in both
\Halpha\ samples, also at similar locations along field-filled lanes
in the two panels. 

The second column shows outer-wing \NaIDone\ and \MgIbtwo\ images of
the same scene at the two sample times.
Each is the starting wavelength of the \acp{CRISP} scan.
For \NaIDone\ this is in the continuum.
For \MgIbtwo, which has extended wings
wider than the \acp{CRISP} prefilter passband and also deep blends, it
is already at 26\% of the line depth.

\begin{figure*}
  \centerline{\includegraphics[width=0.8\textwidth]{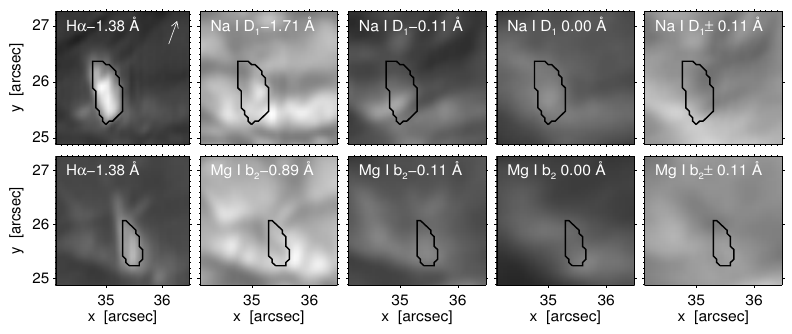}}
  \caption[]{\label{fig:tiny-b} 
  \rev{Magnification of the \acp{EB}s \revtwo{above B} in
  Figure~\ref{fig:cut} in the format of Figure~\ref{fig:tiny}.}
  }
\end{figure*}

The sharpness of the granulation and of numerous intergranular
magnetic bright points in the second panel illustrates the superior
image quality of the \acp{SST}. 
The \acp{EB}s in the first panel are transparent in the second (shown
at larger magnification in 
Figures~\ref{fig:tiny} and \ref{fig:tiny-b}), 
in agreement with
Ellerman's statement that \acp{EB}s do not brighten the continuum.
The granulation scene in the lower panel of the second column differs
not only from \rev{pattern} evolution during the eight-minutes sampling
interval but also from \rev{different sampling through} higher-up
line-wing formation (detailed in
\citeads{2011A&A...531A..17R}). 

Comparison with the \Halpha\ blue-wing granulation in the first column
illustrates the substantial ``flattening'' of granulation in the
latter diagnostic
(\citeads{2005ESASP.596E..15L}), 
which contributes to the relatively high contrast of intergranular
magnetic concentrations in the blue \Halpha\ wing
(\citeads{2006A&A...449.1209L}), 
although they remain less bright than \acp{EB}s (\PaperI, \PaperII).

The third and fourth columns sample the \NaIDone\ and \MgIbtwo\ lines
in their inner blue wings and at their centers. 
These images are less sharp than the far wing ones (also of \Halpha)
because the latter are formed in \acp{LTE} whereas the former belong to the
profile part where scattering dominates heavily in each line,
particularly at its center
(\citeads{2011A&A...531A..17R}). 

Because some contrast features in the inner wings are likely not
caused by profile raising or lowering but by core Dopplershift, we add
Dopplergrams at this sampling wavelength in the final column.
They are defined as blue-wing intensity minus red-wing intensity
normalized by their sum; redshift of the absorption core then produces
enhanced brightness and also brightens line-center samples. 

The \acp{EB}s do not stand out in these Dopplergrams, but other small
bright inner-wing features do.
An example is the bright feature at $(x,y)\is(43,17)$ in both rows.
It is not an \acp{EB}, but a core-redshift or a core plus blue-wing
emission feature or both acting together as the shock grain in the
first quartet of Figure~9 of
\citetads{2010ApJ...709.1362L}. 

These various samples of \NaIDone\ (upper row) and \MgIbtwo\ (lower
row) represent our high-resolution test of Ellerman's statement that
his bombs do not show up in these lines.
Indeed, although there are small bright features in these panels,
there is obviously no direct 1:1 correspondence with the \Halpha\
\acp{EB}s at left. 
None of the latter would be identified as an \acp{EB} from the other
images.
The closest comes the bright \acp{EB} pair \revtwo{below A}, 
which seems to show at least some corresponding brightness in the two
$-0.11$\,\AA\ panels, without Doppler signature.

In Figures~\ref{fig:tiny} and \ref{fig:spectra} we examine \revtwo{the
\acp{EB} pair below A} in detail.
Figure~\ref{fig:tiny} magnifies them in the format of
Figure~\ref{fig:cut}.
The second panel confirms that the first \acp{EB} is transparent in
the continuum. 
The third column does indeed suggest some morphological similarity
between the lower part of the \Halpha\ \acp{EB} and slight brightness
enhancements in the $-0.11$\,\AA\ samplings, especially in \MgIbtwo\
for the second \acp{EB}, but weaker than in the \Halpha-wing.
Other nearby features are as bright. 

The line centers (fourth column) show no clear  signature of the upper
parts of the \acp{EB}s although their height is one Mm or more
(projected lengths in the first column) so that their top reaches
higher than the normal formation height of the \NaIDone\ and \MgIbtwo\
cores (\citeads{2010ApJ...709.1362L}; 
\citeads{2011A&A...531A..17R}). 
\rev{The lower parts of \acp{EB}s may be shielded by 
adjacent undisturbed gas producing normal line cores in slanted
viewing, but their tops should jut out beyond this layer
and be visible if enhanced in these lines.}
This is also evident in the IRIS spectra in \PaperIII\ in which
slanted view lines toward lower \acp{EB} parts show deep \MnI\ blends
on \MgIIhk\ from adjacent undisturbed upper-photosphere gas while
these blends vanish in upper-part spectra.
The conclusion is that the lower parts of these two \acp{EB}s may show
brightening in \NaIDone\ and \rev{especially} \MgIbtwo, but that the
upper parts do not.

The final column shows no pronounced Dopplershifts, whereas in
sampling an \acp{EB} one expects redshift (dark for a bright feature)
for the lower part, blueshift for the upper part (\PaperIII) .

Figure~\ref{fig:spectra} completes our detailed examination of these
two \acp{EB}s by showing line profiles for the lower and upper
samplings marked by black pixels in Figure~\ref{fig:tiny}.
They confirm that only the lower \acp{EB} parts give noticeable wing
enhancements in \NaIDone\ and \MgIbtwo\ but that these remain smaller
than for \Halpha.
The overlying fibril that redshifted the \Halpha\ core did not
redshift the other line cores, but one might speculate that the single
blue-side peak in \MgIbtwo\ from the upper part testifies to an upward
\acp{EB} jet (dotted profile). 

\begin{figure}
  \centerline{\includegraphics[width=0.8\columnwidth]{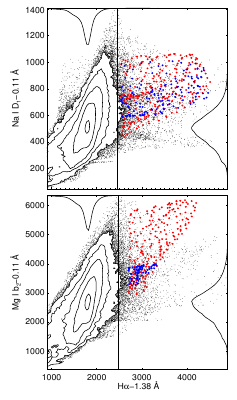}}
  \caption[]{\label{fig:scatter} 
  \rev{Scatter diagrams of \NaIDone\ ({\em upper panel\/}) and
  \MgIbtwo\ \ ({\em lower panel\/}) wing intensities against \Halpha\
  wing intensity for the full field of view (Figure~\ref{fig:full}).
  To avoid plot saturation logarithmic-spaced sample density contours
  are plotted instead of individual pixel samples at large density.
  The histograms along the axes show normalized intensity
  distributions.
  The vertical lines mark 140\% of the mean \Halpha\ value.
  {\em Red pixels\/}: within the \acp{EB} contours in
  Figure~\ref{fig:tiny}.
  {\em Blue pixels\/}: within the \acp{EB} contours in
  Figure~\ref{fig:tiny-b}.
  }}
\end{figure}

We have examined all other \acp{EB}s in Figure~\ref{fig:full} in
comparable detail by using \acp{CRISPEX} to inspect and blink our
various diagnostics at such high magnification, but found no other
with better correspondence that the pair detailed above.
Generally they exhibit brightening in the inner wings of \MgIbtwo, but
not more than in many other features due to network, reversed
granulation or shocks (see
\citeads{2011A&A...531A..17R}). 
\NaIDone\ shows less \acp{EB} brightening or none.
The \MgIbtwo\ profiles display core asymmetry as for the upper pixel
in Figure~\ref{fig:spectra} only rarely.
Thus, while Figures~\ref{fig:tiny} and \ref{fig:spectra} suggest
\rev{some} \acp{EB} signature especially in \MgIbtwo, this is an
exceptional case and it is only a poor case at that.  

We selected \revtwo{the \acp{EB}s below A in Figure~\ref{fig:cut} and
detailed these in Figures~\ref{fig:tiny} and Figure~\ref{fig:spectra}} as
representing a best effort in trying to counter Ellerman's claim.
Its selection as such is illustrated by Figure~\ref{fig:tiny-b}
which repeats the magnified format of Figure~\ref{fig:tiny} for the
\acp{EB}s \revtwo{above B} in Figure~\ref{fig:cut}.
In both \Halpha\ images these are also unmistakable \acp{EB} flames,
but they leave no signature in the other panels of
Figure~\ref{fig:tiny-b}.

Furthermore, we demonstrate the selection of the \acp{EB}s in
Figure~\ref{fig:tiny} as the best counter-example with the full-field
scatter diagrams in Figure~\ref{fig:scatter}. 
The vertical line is one of the criteria defined in \PaperII\ for
automated \acp{EB} detection.
The mountain ridges to the left of these thresholds portray fairly
good correlation between the granulation and network scenes, whereas
the relatively few \acp{EB} pixels lie to their right.
In the upper panel these show no correlation between high \Halpha-wing
intensity and high \NaIDone-wing intensity, but in the lower panel the
red pixels do show such bright--bright correlation for \MgIbtwo.
These are the ones within the contour in the lower panels of
Figure~\ref{fig:tiny}. \revpartwo 
Thus, also in this statistical measure our selected \acp{EB} had
significant concurrent brightening in \MgIbtwo, but it was indeed
exceptional.
The blue pixels for the contours in Figure~\ref{fig:tiny-b} show no
such correlation.

The upshot is that we confirm Ellerman's claim---nearly a
century later.

\begin{figure}
  \centerline{\includegraphics[width=0.93\columnwidth]{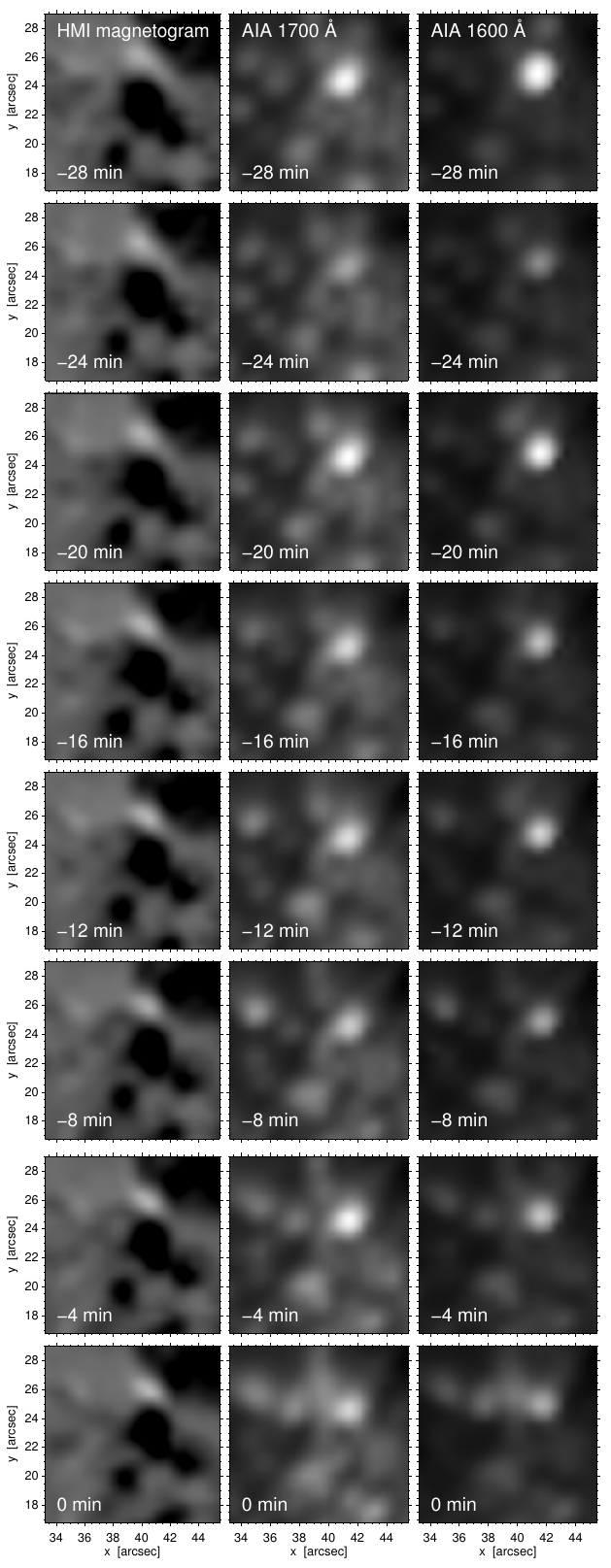}}
  \caption[]{\label{fig:sdo} 
  Time evolution prior to the image samples in
  Figures~\ref{fig:full}--\ref{fig:spectra}.
  {\em Left to right\/}: \acp{HMI} magnetograms, \acp{AIA}\,1700
  and 1600\,\AA\ images. 
  For each the square root of the signal is shown. 
  In addition, the magnetograms were clipped which affects only the
  black polarity. 
  The bytescaling remains the same along columns.
  The bottom row is for 11:29:45\,UT, a few seconds after the
  lower-row samplings in
  Figures~\ref{fig:cut}--\ref{fig:spectra}.
  The other rows span the prior evolution at 4\,minutes intervals;
  $\Delta t \is -8$\,minutes corresponds to the sampling time of
  Figure~\ref{fig:full} and the upper rows of
  Figures~\ref{fig:cut}--\ref{fig:spectra}.
  }
\end{figure}

\paragraph{Hot EB aftermaths}
When we viewed our data as movies we noted, notwithstanding the
bad-seeing moments, a general presence of diffuse bright clouds in
the line-center \NaIDone\ and \MgIbtwo\ sequences near locations where
\Halpha\ \acp{EB}s went off.  
Our impression became that such areas are marked subsequently by
diffuse line-center brightness.

In the fourth column of Figure~\ref{fig:cut} such features are present
above the pore and roughly mimic the arch spanned by the actual
\acp{EB}s in the \Halpha\ panels (first column). 
If one compares the latter with all other panels and asks oneself
whether \acp{EB}s leave some signature in these two lines, these
diffuse line-center clouds seem the most viable candidate.
However, other diffuse brightness patches lower in the panels of
Figure~\ref{fig:cut} do not correspond to \acp{EB}s.

In \PaperIII\ we found that during the aftermath of regular \Halpha\
\acp{EB}s there may appear features with highly enhanced emission in
and very wide profiles of the ultraviolet \SiIV\ and \CII\ lines
sampled by IRIS,
and that these preferentially appear at sites with prolonged repeated
\acp{EB} activity. 

In view of this parallel evidence for \acp{EB} aftermaths \rev{we
suggest tentatively that prior \acp{EB} activity may have caused
diffuse brightness in \NaIDone\ and \MgIbtwo\ and therefore}
turn to the {\em \acp{SDO}} image sequences in order to inspect the
evolution prior to our \acp{SST} snapshots.
This is feasible because \acp{AIA}\,1700 and 1600\,\AA\ images
also display \acp{EB}s, at least the stronger ones (\PaperII\ and
Figure~\ref{fig:full}).
Although they do so less sharply than \acp{CRISP} in \Halpha\ when
there is good seeing at La Palma, they do so everywhere on the solar
disk, all the time (24/7), at good cadence (24\,s), and readily
available in the splendid {\em \acp{SDO}} community service.

Figure~\ref{fig:sdo} shows a time sequence of cutouts of \acp{HMI}
magnetograms and \acp{AIA}\,1700 and 1600\,\AA\ images for the
subfield defined in Figure~\ref{fig:full} and shown in
Figure~\ref{fig:cut}. 
It shows that the bright \acp{EB} in Figure~\ref{fig:cut}, or rather
its location, often harbored a very bright \acp{EB} also before the
\acp{SST} samplings.
Inspection of the earlier \acp{AIA} data showed that this \acp{EB}
flaring actually started already an hour before and that it peaked at
$\Delta t \is -40$~minutes.

In particular, Figure~\ref{fig:sdo} demonstrates that there were also
bright \acp{EB}s at this location during the dozen minutes preceding
our first \acp{SST} sampling and between our two samplings.
The weaker \acp{EB}s to the left of the brightest one in the \Halpha\
panels of Figure~\ref{fig:cut} were also present at
$\Delta t \is -4$~minutes and may have contributed to the arc-shaped
bright cloud in the \MgIbtwo\ line-center panel of
Figure~\ref{fig:cut}.
This prior activity suggests that the diffuse cloud at the centers of
\NaIDone\ and \MgIbtwo\ in Figure~\ref{fig:cut} may represent hot
\acp{EB} aftermaths analogous to the ones found in the \acp{IRIS}
spectra in \PaperIII.

We added the 1600\,\AA\ column to Figure~\ref{fig:sdo} in order to
exclude \acp{FAF}s as potential cause of such diffuse clouds because
\acp{FAF}s tend to have more dramatic hot aftermaths than \acp{EB}s
(\PaperIII). 
However, the scenes in 1700 and 1600\,\AA\ in Figure~\ref{fig:sdo} are
nearly identical, apart from the larger \acp{EB} contrast at
1600\,\AA\ (darkening the background in our column-wise bytescaling).
The prior brightenings also maintained characteristic roundish
\acp{EB} appearance and stability. 
Thus, there were no \acp{FAF}s in this area during this period, and so
the diffuse clouds were not a \acp{FAF} product.

The sequence of \acp{HMI} magnetograms in the first column suggests
that the major patch of white polarity diminished on its lower-right
side where it was adjacent to the larger and stronger black patch and
where the brightest \acp{EB}s occurred. 
There were weaker-field encounters at the weaker \acp{EB} sites.
The resolution and sensitivity are insufficient to establish
small-scale field cancelation, but these patterns suggest that it may
well have happened.

\section{Conclusions}\label{sec:conclusion}

Figure~\ref{fig:cut} confirms Ellerman's statement that his bombs do
not brighten the \NaID\ and \MgIb\ lines.  
Figures~\ref{fig:tiny} and \ref{fig:spectra} do suggest \acp{EB}-foot
brightening in their wings, but not much relative to other nearby
features. 
Also, this is the best case; the other \acp{EB}s in our samples show
less correspondence.

However, at the line centers there are diffuse bright clouds that may
represent \acp{EB} aftermaths comparable to those diagnosed from
\acp{IRIS} spectra in \PaperIII.

The interpretative suggestion from these results and from our results
in \PaperIII\ is obvious: the upper parts of \acp{EB}s are too hot to
radiate in these lines, likely from neutral-species ionization, and
hot enough to radiate in \SiIV\ lines.
Much higher temperatures seem involved than in all \acp{EB} modeling
so far. 

\rev{In addition, we speculate that} in \acp{EB} aftermaths the
neutral species may reappear through recombination in cooling
post-bomb ``mushroom'' clouds. 
For these, the observational suggestion is also obvious: perform
similar dual-line imaging spectroscopy as presented here but over
longer duration and together with \acp{IRIS} spectrometry, emphasizing
\acp{EB} and \acp{FAF} aftermaths.

The challenge for \acp{EB} modeling is to meet the wide set of diverse
constraints posed by our high-resolution observations in \Halpha\
(\PaperI\ and \PaperII), the ultraviolet IRIS lines (\PaperIII), and
the \NaIDone\ and \MgIbtwo\ lines presented here.
\rev{Modeling that explains all of these may also give insight into
the nature of aftermath clouds.}

\begin{acknowledgements}
This research was supported by the Research Council of Norway and by
the European Research Council under the European Union's Seventh
Framework Programme (FP7/2007-2013)/ERC Grant agreement nr.\ 291058.
The \acp{SST} is operated by the Institute for Solar
Physics of Stockholm University in the Spanish Observatorio del Roque
de los Muchachos of the Instituto de Astrof{\'\i}sica de Canarias on
the island of La Palma.
\revpar
\end{acknowledgements}


\end{document}